# Superconductivity of metastable dihydrides at ambient pressure

Heejung Kim[1], Ina Park[2], J. H. Shim[2], D. Y. Kim[3,4*`]

[1]*Department MPPHC-CPM, Max Plank POSTECH/Korea Research Initiative, Pohang 37673, Korea*

[2]*Department of Chemistry, POSTECH, Pohang 37673, Korea*

[3]*Center for High Pressure Science and Technology Advanced Research, Shanghai 201203, China*

[4]*Division of Advanced Nuclear Engineering, POSTECH, Pohang 37673, Korea*

**ABSTRACT**

Hydrogen in metals is a significant research area with far-reaching implications, encompassing diverse fields such as hydrogen storage, metal-insulator transitions, and the recently emerging phenomenon of room-temperature ($T_C$) superconductivity under high pressure. Hydrogen atoms pose challenges in experiments as they are nearly invisible, and they are considered within ideal crystalline structures in theoretical predictions, which hampers research on the formation of meta-stable hydrides. Here, we propose pressure-induced hydrogen migration from tetrahedral site ($T$) to octahedral site ($O$), forming $LaH_x^O H_{2-x}^T$ in cubic $LaH_2$. Under decompression, it retains $H_x^O$ occupancy, and is dynamically stable even at ambient pressure, enabling a synthesis route of metastable dihydrides via compression-decompression process. We predict that the electron phonon coupling strength of $LaH_x^O H_{2-x}^T$ is enhanced with increasing $x$, and the associated $T_C$ reaches up to 10.8 K at ambient pressure. Furthermore, we calculated stoichiometric hydrogen migration threshold pressure ($Pc$) for various lanthanides dihydrides ($RH_2$, where $R$=Y, Sc, Nd, and Lu), and found an inversely linear relation between $Pc$ and ionic radii of $R$. We propose that the highest Tc in the face-centered-cubic dihydride system can be realized by optimizing the $O/T$-site occupancies.

**INTRODUCTION**

Since Ashcroft suggested that hydrogen-dense materials could have high superconducting $T_C$ at relatively lower pressure than pure solid hydrogen by chemical precompression onto the hydrogen sublattice[1], the hydride researches have been intensively studied both in experiments and theories[2–8]. The remarkable advances of accessible pressures in diamond anvil cell and the development of theoretical methods have accelerated the researches on hydride superconductor. Among them, the superconductivity in sulfur hydride was observed with maximum $T_C$ of 203 K[9]. Most of all, superhydride materials[10–12] in the form of symmetric clathrate structures are remarkable propose to rise the superconducting $T_C$ by enhancing electronic density of states by hydrogens at the Fermi level. Eventually, experimental studies showed that $LaH_{10}$ exhibits high-$T_C$ superconductivity with $T_C$ = 250 K at 300 GPa[13,14].

Despite these observations of superconductivity in rare-earth superhydrides at extremely high-pressure, the stabilization at ambient pressure is marginal due to a strong Peierls distortion leading to the decomposition of the hydrogen sublattice into hydrogen molecules. Recently, there is a strong research interest in finding ways to lower the transition pressure, making it more practical and feasible for technology. For examples, meta-stable superhydrides derived from $LaH_{10}$ have been proposed as candidates exhibiting superconductivity at lower pressures. Cataldo *et al.* proposed a ternary sodalite-like $LaBH_8$ consisting of La-B scaffold and hydrogen sublattice[15]. They predicted that $LaBH_8$ is thermodynamically stable above 100 GPa, and dynamically stable down to 40 GPa with $T_C$ of 126 K. Zhang *et al.* suggested ternary hydrides with a fluorite-type backbone ($XH_8$)[16]. They reported that the $LaBeH_8$ is thermodynamically stable above 98 GPa with $T_C$ of 192 K, and meta-stable under a pressure of 29 GPa with $T_C$ of 193K.

Rare-earth dihydride $R$H$_2$ is the ground state at ambient pressure, which is metallic. $R$H$_2$ consists of face-centered-cubic (*fcc*)-$R$ with two hydrogen atoms intercalated at symmetric tetrahedral (*T*-) sites keeping the octahedral (*O*-) site empty[17] (see the Fig. 1 (a)). When additional hydrogen atoms are added to $R$H$_2$, they locate at *O*-site forming $R$H$_3$. It is known that the $T_C$s in $R$H$_2$ are negligible at ambient pressure[18–21].

Machida *et al.* reported that LaD$_2$ undergoes disproportionation reaction to form both of the D-deficient NaCl-type LaD and D-rich LaD$_3$ above 11 GPa[22]. Furthermore, they also found that the LaD is recovered below 4 GPa by decompression processes,[23] and LaD$_3$ was still observed at the measured lowest pressure (0.2 GPa), revealing irreversibility of pressure-induced hydrogen migration. They mentioned that D atoms of LaD$_2$ in *T*-site are transferred into *O*-site by compression. In experiments, metastable hydrides have partial occupation of the two interstitial sites.

In here, we focus on dynamical stable $R$H$_2$ structure with partial occupation of *T*-site and occupied *O*-site in LaH$_x^O$H$_{2-x}^T$ as an attempt to find superconductivity at ambient pressure. We studied the structural stability and superconductivity of LaH$_x^O$H$_{2-x}^T$, and demonstrated that LaH$_x^O$H$_{2-x}^T$ is dynamically stable at ambient pressure. Also, we obtained that λ of LaH$_x^O$H$_{2-x}^T$ becomes enhanced with increasing the occupation of O-site by H, which yields the increment of $T_C$ up to 10.8 K at ambient pressure. Furthermore, we generalized the idea of superconductivity induced by partial occupancy of H atom to other lanthanides dihydrides ($R$H$_2$, Y, Sc, Nd, and Lu). In case of LuH$_1^O$H$_1^T$, we obtained $T_C$ = 36.23 K with λ = 1.174 at 20 GPa.

# RESULTS

## The metal-stability of LaH$_2$

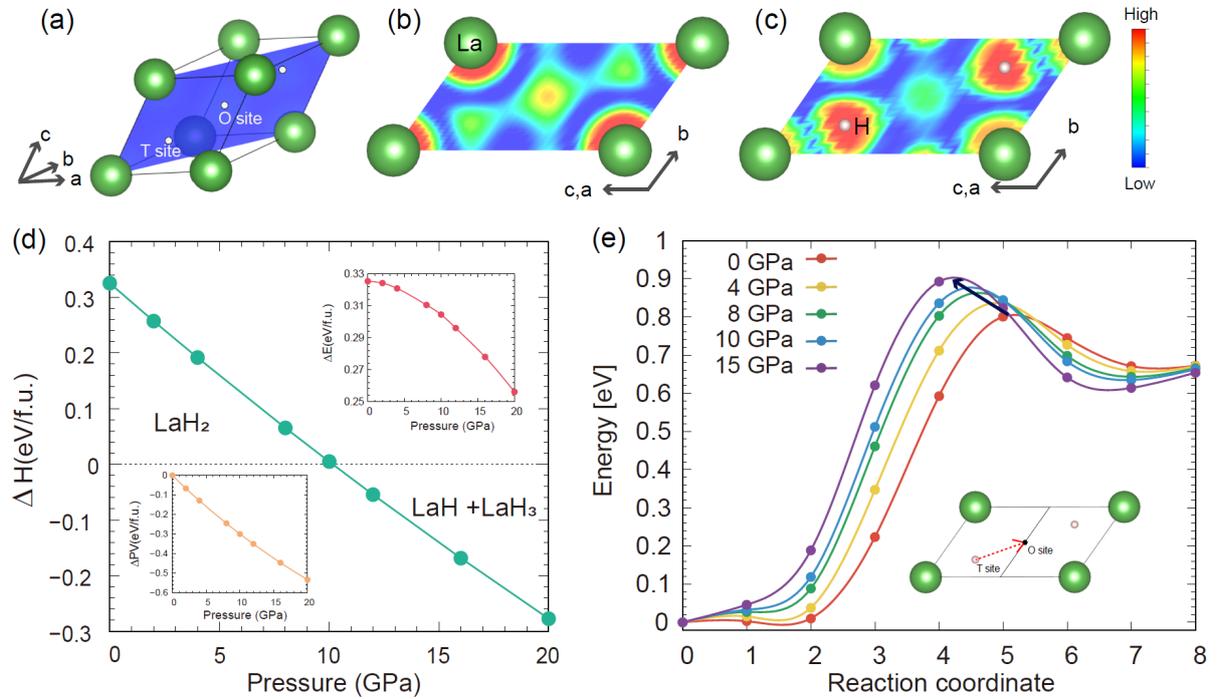

Figure 1. (a) The crystal structure of pure *fcc*-La (space group: $Fm\bar{3}m$) with two tetrahedral sites (*T*) and one octahedral site (*O*) within La lattice. Both *T*- and *O*-sites are on [110] plane in blue color. (b) and (c) Electron localization function of *fcc*-La and LaH$_2$ on [110] plane, respectively. (d) The relative formation enthalpy ΔH of LaH$_2$ and LaH + LaH$_3$ as a function of pressure. The insets show internal energy, ΔE (upper right), and ΔPV (lower left) variation with pressure. (e) The variation of the energy along the reaction path between LaH$_2$ and LaH$_1^O$H$_1^T$ as a function of pressure. (The inset shows the moving path (red dashed arrow) of one hydrogen from *T*-site to *O*-site within primitive cell.) The black arrow indicates the variation of the highest point of kinetic barriers with pressure.

Figure 1 (a) shows a pure face-centered-cubic (*fcc*)-La (space group: $Fm\bar{3}m$). Electron localization function (ELF) of fcc-La in Fig. 1 (b) reveals that the electrons are well localized in the regions corresponding to *T*- and *O*-sites, which is consistent with the previous reports[24,25]. Bader net charge [26] at O-site is estimated to be $0.02e$, exhibiting the non-nuclear attractor (NNA) formation as an electride, possibly hosting hydrogen atoms. La atoms forms strong ionic bonding with H atoms at T-site, and NNA captures hydrogen atoms at O-site. Figure 1 (c) shows that NNA still exists at O-sites in $LaH_2$ indicating H atoms can be captured.

We calculated the relative formation enthalpy ($\Delta H$) from $2LaH_2$ to $LaH + LaH_3$ under pressure as shown in Fig. 1(d). $\Delta H$ is obtained from $[H_{LaH} + H_{LaH_3}]/2 - H_{LaH_2}$, where the enthalpy is defined as $H = E_0 + PV$, $E_0$ is internal energy at a given pressure $P$ and volume $V$. At ambient pressure, the $\Delta H$ is approximately 0.32 eV/f.u., indicating stable $LaH_2$ phase. With increasing pressure, $\Delta H$ is decreased due to the rapid change of $\Delta PV$ (see the lower left inset). Eventually, $\Delta H$ turns negative around the transition pressure ($Pc$) at 10 GPa. This result means that the stoichiometric disproportionation reaction from $2LaH_2$ to $LaH$ and $LaH_3$ occurs at the pressure, which is consistent with previous result[22].

To understand the stability of H occupation at *O*-site within $LaH_2$, we conducted nudged elastic band (NEB) calculations migrating one H atom located *T*-site (¼ , ¼, ¼) into *O*-site (½, ½ , ½) in $LaH_1^O H_1^T$ as shown in Fig.1(e). $LaH_2^T$ is definitely stable against $LaH_1^O H_1^T$ configuration by 0.67 eV/f.u. at ambient pressure. However, one can easily find that there is a kinetic barrier between T- and O- sites, suggesting $LaH_1^O H_1^T$ can be a metastable phase. The peak of kinetic barrier shifts to higher with pressure (i. e. the black arrow in Fig 1 (e)), further stabilizing $LaH_1^O H_1^T$ configuration. We also tested it in a supercell (see Supplementary material Figure S1) to check the size effect and it shows consistent result.

Interestingly, it was experimentally reported that the pressure-induced disproportional reaction from LaD$_2$ to LaD + LaD$_3$ is irreversible under decompression[22], which can be explained by our calculations. Thus, compression-decompression process can open a pathway to study metastable LaH$_2$ with partial O-site occupancy at ambient pressure.

**Electronic structure**

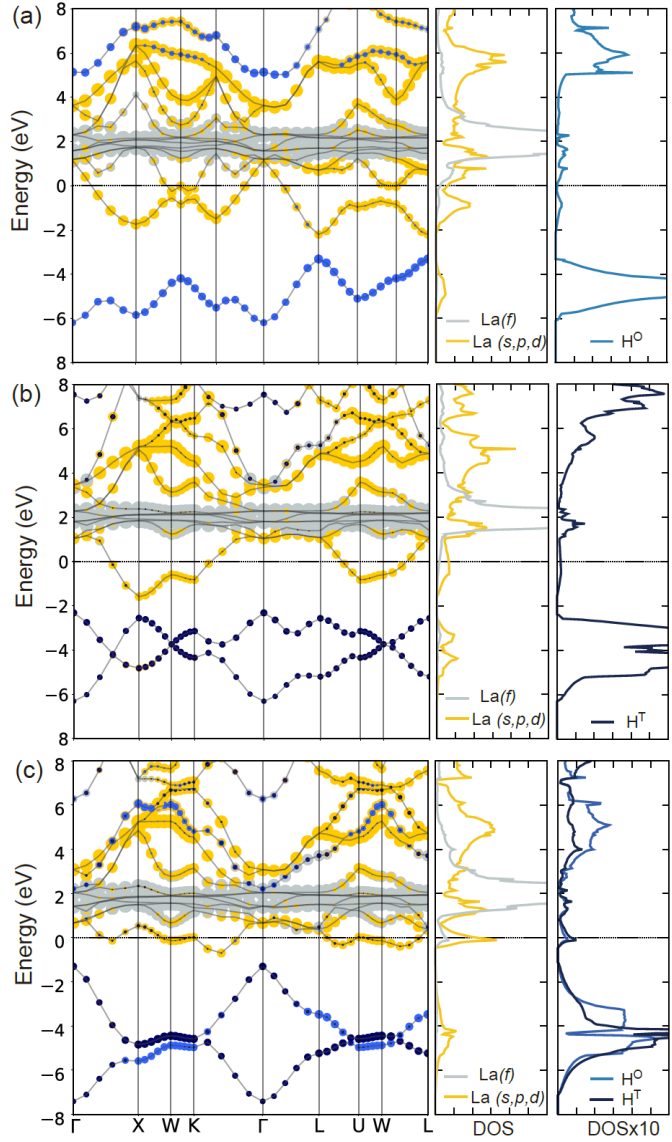

Figure 2. (a)-(c) Band structures and DOS of LaH, LaH$_2^T$, and LaH$_1^O$H$_1^T$ with fat bands along symmetry lines of the *fcc*-BZ. The sizes of circles in the band structures indicate the contribution of each orbital.

Figure 2 shows the electronic bands and density of state (DOS) along the high-symmetry k points in the *fcc*-Brillouin zone (BZ) for LaH, LaH$_2^T$ and LaH$_1^O$H$_1^T$. In all cases, the energy levels near the Fermi level ($E_F$) are mainly contributed by *spd*-bands of La, and *f*-bands of La are located well above the $E_F$. The *s* band of H in LaH$_1^O$H$_1^T$ below -2 eV is more dispersive than these of LaH and LaH$_2^T$ as shown in Fig. 2 (c) because its H-H distance is shorter than others (LaH: 2.72Å, LaH$_2^T$: 2.80Å, LaH$_1^O$H$_1^T$: 2.37Å). The DOS at $E_F$ in LaH$_1^O$H$_1^T$ is much higher than LaH and LaH$_2^T$ due to a strong hybridization between H and La bands, which also can be explained by the flat band around W point in the BZ. Thus, metastable dihydride possesses much enhanced charge carriers.

**Superconducting property**

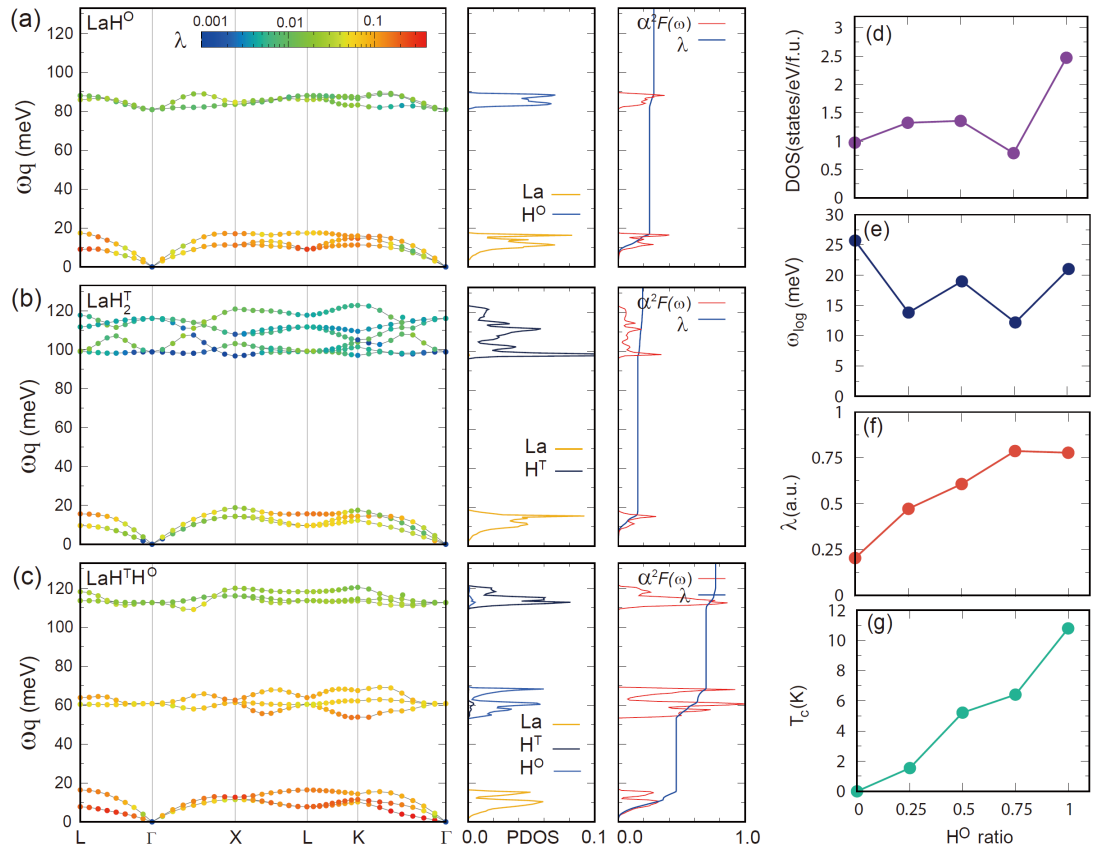

Figure 3. The phonon dispersion, PDOS, and $\alpha^2F(\omega)$ of (a) LaH$_1^O$, (b) LaH$_2^T$ and (c) LaH$_1^O$H$_1^T$. The color code in the phonon dispersion represents the magnitude of $\lambda_{qv}$. The values of (d) the total DOS at $E_F$, (e) $\omega_{log}$, (f) $\lambda_{tot}$, and (g) superconductivity $T_C$ as the function of $x$ in $LaH_x^O H_{2-x}^T$.

In order to investigate the structural stability, electron-phonon coupling strength, and superconductivity $T_C$, we calculated the phonon dispersion, phonon density of states (PDOS), Eliashberg spectral functions [$\alpha^2 F(\omega)$], and the electron phonon coupling constants ($\lambda_{qv}$) of $LaH_1^O$, $LaH_2^T$ and $LaH_1^O H_1^T$ at ambient pressure as shown in Figs. 3 (a) ~ (c). In all cases, the contributions of La atoms dominate mostly in the low-frequency region (0 ~ 5 meV) while H atoms contribute to the high frequency region. The phonon frequencies of H atoms in $LaH_2^T$ (~110 meV) are higher than that of $LaH_1^O$ (~90 meV) due to the shorter bonding distances between La and H. In $LaH_1^O$ and $LaH_2^T$, the contributions to spectral function $\alpha^2 F(\omega)$ are from both La and H, but their electron phonon couplings ($\lambda_{tot}$) are mainly contributed by La due to the small DOS of H atoms at $E_F$ as shown in Figs. 2 (a) and (b). Accordingly, their total $\lambda_{tot}$s are quite small (0.28 for $LaH_1^O$ and 0.20 for $LaH_2^T$), which results in negligibly small $T_C$ ~ 0 K for both materials.

Interestingly, $LaH_1^O H_1^T$ shows a quite different phonon dispersion behaviors compared with $LaH_1^O$ and $LaH_2^T$. The phonon dispersion and the PDOS reveals a coupling between $H^T$ and $H^O$ manifesting the peaks near 60 meV. At these ranges, $\alpha^2 F(\omega)$ and $\lambda$ are enhanced due to the increase of DOS at $E_F$. Furthermore, the large DOS of La at $E_F$ increases of $\alpha^2 F(\omega)$ and $\lambda$ near low frequencies. Accordingly, the $T_C$ of $LaH_1^O H_1^T$ exhibits 10.8 K with $\lambda_{tot}$ = 0.778. This study highlights that the metastable structure of $LaH_2$, with a change in the site occupancy of $O$-site/$T$-site, can increase $T_C$ dramatically, providing a mechanism to enhance superconductivity transition temperature at ambient pressure.

We also have calculated the change of superconductivity properties depending on the ratios of $H^O$ and $H^T$ for the given $LaH_2$ stoichiometry. Figure 3 (d~g) present the DOS at $E_F$, the $\lambda_{tot}$, the logarithmic average phonon frequency ($\omega_{log}$), and the $T_C$ as a function of $O$-site hydrogen ratio. The DOS at $E_F$ is amplified with increasing the $O$-site occupation, resulting in an enhancement of $\lambda_{tot}$ as shown in Figs. 3 (d) and (f), providing the optimal $H^O/H^T$ ratio for the highest $T_C$ in $LaH_x^O H_{2-x}^T$. Since the $T_C$ is obtained from McMillan equation[27,28]: $\{\omega_{log}/1.2\}\{\exp[-1.04(1+\lambda)/\lambda(1-0.62\mu^*)-\mu^*]\}$, it depends on both $\omega_{log}$ and $\lambda_{tot}$. Even though $\omega_{log}$ fluctuates with the increased $O$-site occupation in Fig. 3 (e), $T_C$s are monotonically enhanced mainly due to the exponential dependence on $\lambda_{tot}$ as $x$ approaching from 0 to 1 as shown in Fig. 3 (g). Thus, we conclude that $T_C$ has a maximum value when the occupation of $H^O$ and $H^T$ are evenly distributed.

**Generalized meta-stable $R$H$_2$**

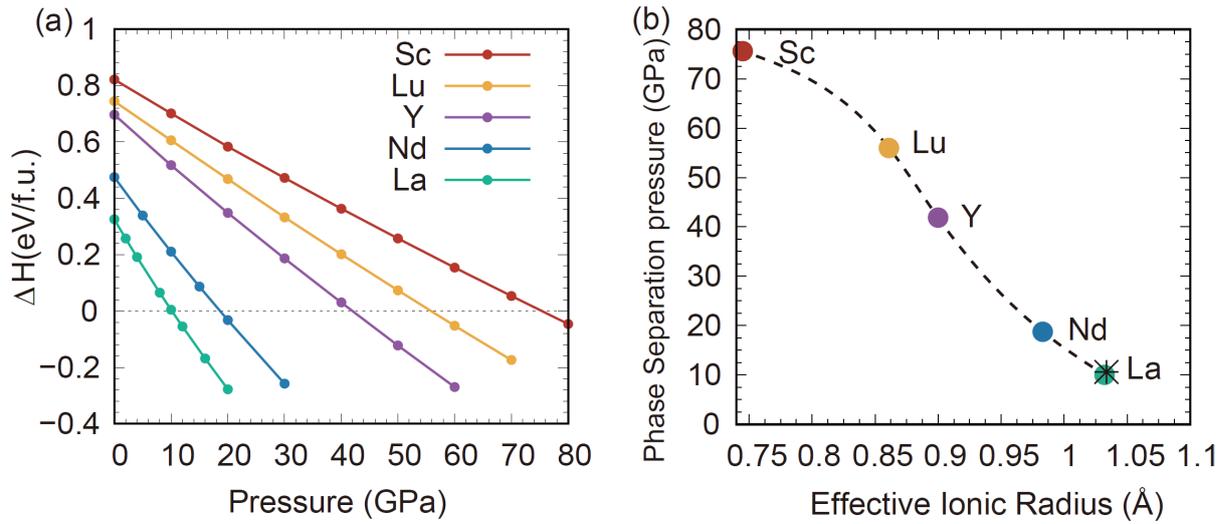

Figure 4. (a) The relative formation enthalpy of $\frac{1}{2}(RH + RH_3)$ with respect to the $RH_2$ by changing pressure. (b) The critical pressure required for the disproportionation reaction into $RH + RH_3$ with respect to the ionic size. (experimental result [22] for La in the star)

It is also straightforward to expect that this mechanism could also be applicable to other materials with the same structural frame of rare earth hydrides such as *fcc-R*H$_2$ ($R$: La, Nd, Y, Lu, and Sc). Figure 4 (a) shows relative formation enthalpies of $R$H$_2$ as a function of pressure.

The disproportional reaction for stoichiometric change from *T*-site to *O*-site ($RH_2 \rightarrow RH + RH_3$) happens at ~ 10 GPa for La and ~75 GPa for Sc. In Fig. 4 (b), we plotted *Pc* vs effective ionic radius of studies compounds and interestingly found a linear relation: the smaller the ionic size, the higher the *Pc* required. In LaH$_2$, our result is in a good match with experimental data (see the star in Fig. 4(b)). Thus, Fig. 4 (b) can be a guideline for experimental synthesis of metastable dihydrides superconductors.

Very recently, Nathan *et al.* reported that LuH$_{3-\delta}$N$_\delta$ ($Fm3m$), one of lanthanide hydrides, exhibits maximum $T_C$ of 296 K at 1 GPa, which nearly reaches the room temperature superconductivity at ambient conditions[29]. It is quite striking because the result is against our conventional wisdom about high-Tc superhydrides and it is not consistent with previous experimental data for LuH$_3$ reporting only 12.4 K at 122 GPa[30]. We need to note that this discovery[29] hasn't been reproduced by other groups yet, and several follow-up studies claim absence of superconductivity[31,32]. The reported process of synthesizing LuH$_{3-\delta}$N$_\delta$ seems to depend on pressure and temperature, possibly forming metastable hydrides. We have calculated the superconducting properties of lutetium hydride (LuH$_1^O$H$_1^T$) as shown in Fig. S3. At ambient pressure, LuH$_1^O$H$_1^T$ shows phonon instability (imaginary phonon frequency) as shown in S3 (a). Pc for LuH$_1^O$H$_1^T$ is predicted to be ~ 55 GPa and thus the stable phonon can be obtained at relatively high pressure. As shown in Fig. S3 (b), our result shows that LuH$_1^O$H$_1^T$ becomes stable at pressures above 10 GPa. Also, we obtained that $T_C$ of LuH$_1^O$H$_1^T$ is 36.23 K with $\lambda = 1.174$ while $T_C$ of LuH$_2^T$ is 0.09 with $\lambda = 0.29$, exhibiting nontrivial enhancement of electron-phonon coupling. Nitrogen-doping may increase the Tc by shifting $E_F$, however, we cannot find any clue to reach the recently observed room temperature superconductivity.

## CONCLUSIONS

In summary, we studied hydrogen migration between *T*- and *O*- sites in LaH$_2$. Pressure stabilizes *O*-site and eventually disproportional reaction happens at the threshold pressure (*Pc*), which induces migration in hydrogen positions from LaH$_2^T$ at ambient pressure to LaH$_x^O$H$_{2-x}^T$ at high pressure. Lanthanum dihydride with partial hydrogen atoms at octahedral sites (LaH$_x^O$H$_{2-x}^T$) is dynamically stable at ambient pressure and it possesses enhanced electron-phonon coupling strength to exhibit finite superconductivity transition temperature. We also generalized to other lanthanides hydrides (*R*H$_2$, where *R*=Y, Sc, Nd, and Lu) and found that they follow the same trend as LaH$_2$. Moreover, we found a simple connectivity between Pc and ionic radius of lanthanides, providing a guideline to further experiments. We also obtained that LuH$_1^O$H$_1^T$ has Tc = 36.23 K with $\lambda = 1.174$ at 20 GPa.

## METHODS

For the band structure calculations, we employed the pseudopotential method in Vienna Ab initio Simulation Package (VASP)[33,34]. The electronic charge density was evaluated up to the kinetic energy cutoff of 350 eV. The face-centered-cubic (fcc) Brillouin-zone (BZ) integration for the self-consistent calculations was carried out with 12 x 12 x 12 *k*-points. The volume optimization was done by using the Hellmann-Feynman force theorem[35]. To calculate the barrier from tetrahedron site of hydrogen to octahedron site, we performed nudged elastic band calculations[36]. We used seven images of the intermediate configurations connecting the two most stable structures. Each image was relaxed until the force on each atom was smaller in magnitude than 0.05 eV/ Å$^{-1}$. Phonon dispersion and electron-phonon coupling constants were obtained by using the density-functional perturbation theory implemented in the Quantum ESPRESSO package[37] with PAW-PBE pseudopotential of PSlibrary[38]. The dynamical matrices were calculated on 8 x 8 x 8 **q**-grid for fcc-structure and 2 x 2 x 2 **q**-grid for conventional cubic structure. All the phonon calculations were performed after the volume relaxation of structure. We checked that two sets of methods (VASP and Quantum ESPRESSO) satisfy the consistency of the band structures.

ACKNOWLEDGEMENTS

This work was supported by the National Research Foundation (NRF) Korea (Grant No. 2021R1C1C2011276, RS-2023-00272090). J. H. S. acknowledges the support by Brain Pool program funded by the Ministry of Science and ICT through the NRF (2022H1D3A2A01096414). D. Y. K. acknowledges the support from National Natural Science Foundation of China (11774015).

AUTHOR CONTRIBUTIONS

H. Kim performed the calculations, D. Y. Kim conceived and supervised the project. All authors contributed to the discussion of the results and participated in preparing the manuscript.

COMPETING INTERESTS

The authors declare no competing interests.

# Superconductivity of metastable dihydrides at ambient pressure

Heejung Kim[1], Ina Park[2], J. H. Shim[2], D. Y. Kim[3,4*]

[1]Department MPPHC-CPM, Max Plank POSTECH/Korea Research Initiative, Pohang 37673, Korea

[2]Department of Chemistry, POSTECH, Pohang 37673, Korea

[3]Center for High Pressure Science and Technology Advanced Research, Shanghai, China

[4]Division of Advanced Nuclear Engineering, POSTECH, Pohang 37673, Korea

Email: duckyoung.kim@hpstar.ac.cn

# The NEB calculation of $La_4H_8$

We have performed NEB calculation for a conventional unit cell (four formula units of LaH$_2$) migrating one H atom from T-site (0.75, 0.25, 0.25) to O-site (0.5, 0.5, 0.5) as shown in Fig. S 1(a). We observed that the dynamical stability in $LaH^O_{0.25}H^T_{1.75}$ is enhanced with the increase of energy barrier under pressure as shown in Fig. S 1(b). We note that at ambient pressure, $LaH^O_{0.25}H^T_{1.75}$ is calculated to show stable phonon dispersion relations while we cannot find the barrier in Fig S1 (b). We speculate that stability of $LaH^O_{0.25}H^T_{1.75}$ may be obtained along other paths.

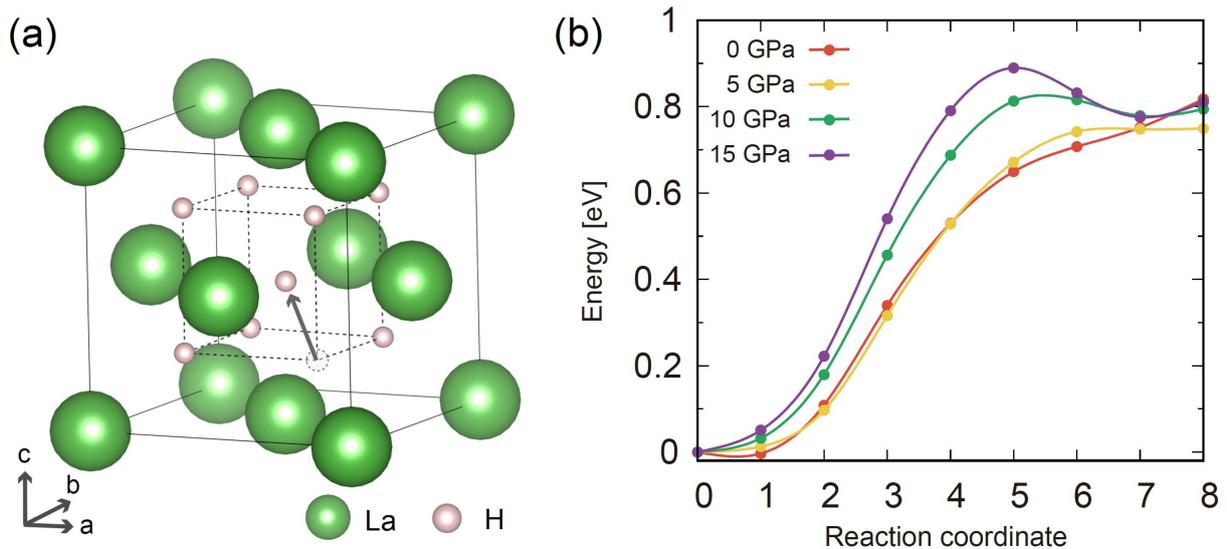

Figure S1. (a) Crystal structure of cubic 4LaH$_2$. The gray arrow shows the moving path of one H atom. (b) Change of the energy as reaction path (see the arrow in (a)) from $4LaH^O_0H^T_2$ to $4LaH^O_{0.25}H^T_{1.75}$ by changing pressure using NEB calculation.

**The phonon calculation of $LaH_x^O H_{2-x}^T$**

Figure S.2 shows phonon dispersion, phonon density of states (PDOS), Eliashberg spectral function [$\alpha^2 F(\omega)$], and the electron phonon coupling constant ($\lambda_{qv}$) in $LaH_{0.25}^O H_{1.75}^T$, $LaH_{0.5}^O H_{1.5}^T$, and $LaH_{0.75}^O H_{1.25}^T$. As $H^O$ is added up $LaH_x^O H_{2-x}^T$, the hybridization of $H^O$ and $H^T$ is increased, which induces the enhancement of total $\lambda_{\text{tot}}$.

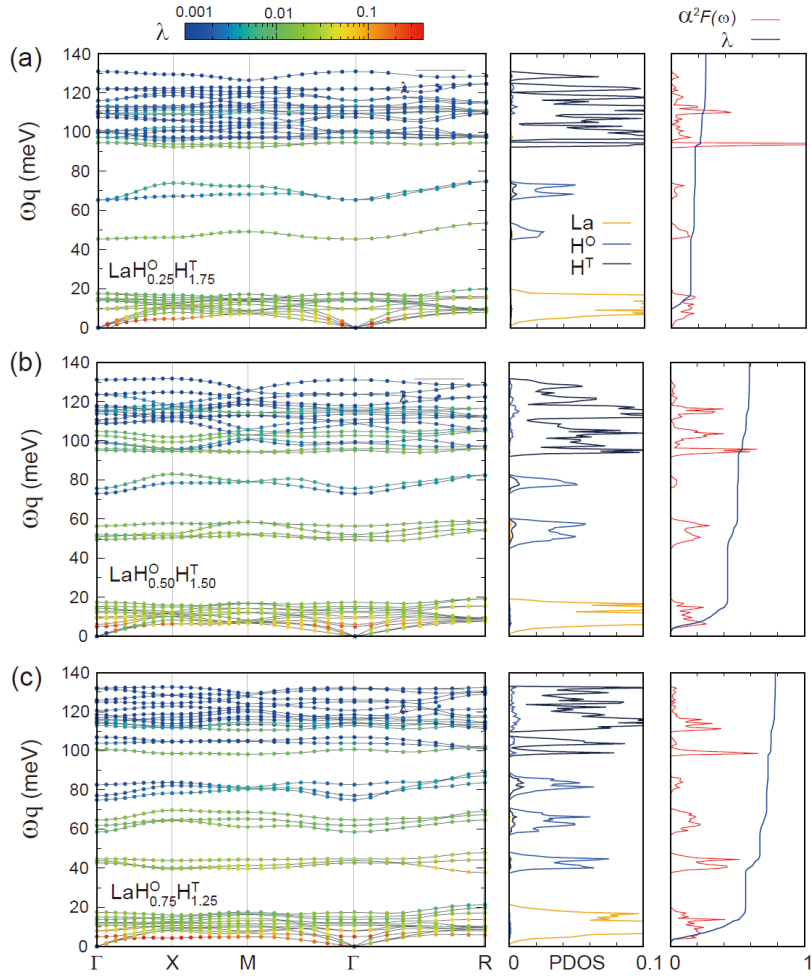

Figure S2. The phonon dispersion with $\lambda_{qv}$, PDOS, and $\alpha^2 F(\omega)$ of (a) $LaH_{0.25}^O H_{0.75}^T$, (b) $LaH_{0.50}^O H_{1.5}^T$, and (c) $LaH_{0.75}^O H_{0.25}^T$.

## The phonon calculation of LuH$_2$ at ambient and 20 GPa

We have calculated phonon dispersion, phonon density of states (PDOS), Eliashberg spectral function [$\alpha^2 F(\omega)$], and the electron phonon coupling constant ($\lambda_{qv}$) of LuH$_1^O$H$_1^T$ at ambient pressure and high pressure. We have found the presence of phonon softening instability (imaginary phonon frequency) indicating structural instability at ambient pressure. As shown in PDOS, the softening arises from $H^O$. However, we have observed the hardening of phonon due to the increment of the bond strength between ions. In practice, we have found that there is no phonon softening stability in 20 GPa, indicating dynamical stability.

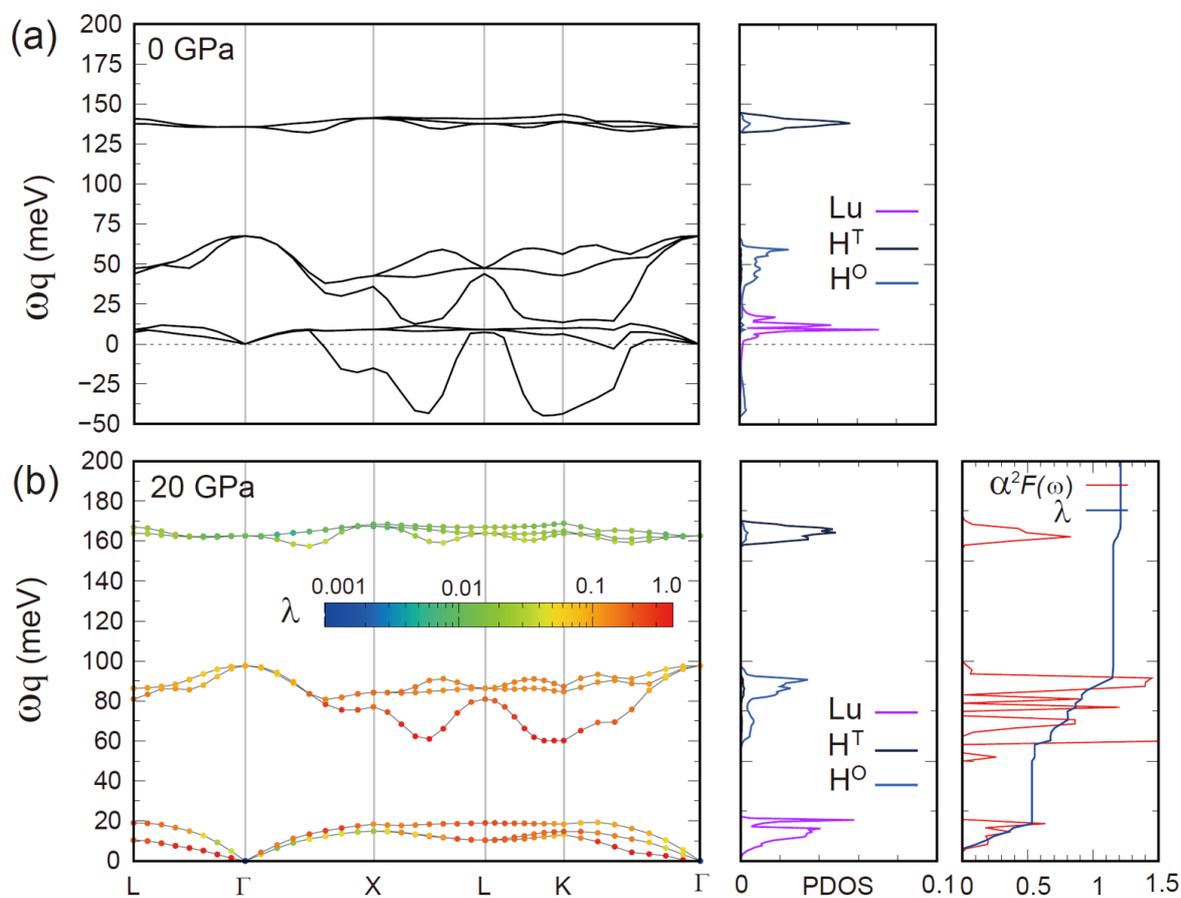

Figure S3. The phonon dispersion relations, PDOS, and $\alpha^2 F(\omega)$ of LuH$_1^O$H$_1^T$ at 0 GPa (a) and 20 GPa (b).